\documentstyle[twoside,fleqn,espcrc2]{article}

\title{\bf Determination of the strange-quark mass from QCD pseudoscalar
sum rules\thanks{Work supported in part by the Volkswagen Foundation}}

\vspace{0.5cm}
\author{\bf C. A. Dominguez, L. Pirovano\address{Institute of Theoretical 
Physics and
Astrophysics\\University of Cape Town, Rondebosch 7700, South Africa} 
and K. Schilcher\address{Institut f\"{u}r Physik, Johannes 
Gutenberg-Universit\"{a}t\\
Staudingerweg 7, D-55099 Mainz, Germany}} 

\begin{document}

\begin{abstract}
\noindent
A new determination of the strange-quark mass is discussed,
based on the two-point function involving the axial-vector current
divergences. This Green function is known in perturbative QCD up to
order $\cal {O}$$(\alpha_s ^{3})$, and up to dimension-six in the non-
perturbative domain. The hadronic spectral function is parametrized
in terms of the kaon pole, followed by its two radial excitations, and
normalized at threshold according to conventional chiral-symmetry. The
result of a Laplace transform QCD sum rule analysis of this two-point
function is: $\bar {m}_{s}(1 GeV^{2}) = 155 \pm 25 MeV$.
\end{abstract}

\maketitle

Past optimistic expectations to determine the values of the light quark
masses with increasing precision, in the framework of QCD sum rules, are
presently being brought into question. This is a result of the uncovering
of various systematic uncertainties, previously unknown or underestimated.
First, there is the problem of reconstructing the hadronic spectral function
using experimental data on the masses and widths of the ground state hadrons
and their (radially) excited states. This information is far from being
sufficient to achieve the task. While narrow resonances may be reasonably
parametrized by Breit-Wigner forms, this is not the case for broad states,
e.g. the $a_{1} (1260)$ axial-vector meson. Still more important is the
potential presence of non-resonant background which could interfere
destructively or constructively with the resonance parametrization. While the
by now standard procedure of normalizing hadronic spectral functions at
threshold using chiral symmetry \cite{CAD1} does provide some form of
non-resonant background, this may not be enough. To illustrate the point,
let us consider the determination of $(m_{u} + m_{d})$ using Finite Energy
QCD sum Rules (FESR) in the pseudoscalar channel, and compare the result of
\cite{CAD2}
\begin{equation}
(m_{u}+m_{d}) \; \mbox{(1 GeV)} = 15.5 \pm \mbox{2.0 MeV} \; ,
\end{equation}
with that of \cite{BIJ1}
\begin{equation}
(m_{u}+m_{d}) \; \mbox{(1 GeV)} = 12.0 \pm \mbox{2.5 MeV} \; .
\end{equation}

While compatible within errors, these two results lead to rather different
values for the strange quark mass. In fact, using the current algebra ratio
\cite{GL}
\begin{equation}
\frac{m_{s}}{m_{u}+m_{d}} = 12.6 \pm 0.5 \; ,
\end{equation}
one finds from Eqs. (1) and (2), respectively
\begin{eqnarray}
m_{s} \; \mbox{(1 GeV)} \; = \; 
\begin{array}{lcl}
195 \pm 28 \; \mbox{MeV}\\ [.3cm]
151 \pm 32 \; \mbox{MeV} \; .
\end{array}
\end{eqnarray}
The problem here is that the same raw data for resonance masses and
widths, plus the same threshold normalization from chiral-perturbation
theory has been used in both analyses \cite{CAD2} and \cite{BIJ1}.
The difference in the results arises mainly from the choice of the functional
form for the hadronic spectral function. Since there is no {\bf direct}
experimental information on this function over a wide range of energies, this
issue remains unresolved. Results (1) and (2) should be considered together,
with the spread in values providing an estimate of systematic uncertainties
from the hadronic sector. A more dramatic illustration of the size of these
uncertainties comes from the QCD Laplace sum rule determination of $m_{s}$
in the scalar channel.  Here, the available experimental data on $K-\pi$
phase shifts \cite{EXP1} should allow, in principle, for a clean
reconstruction of the hadronic spectral function from threshold up to
$s \simeq 7 GeV^{2}$. Assuming that the only non-resonant background is
the one provided by the chiral-symmetry normalization of the hadronic
spectral function at threshold, two independent determinations give
\begin{eqnarray}
m_{s} \; \mbox{(1 GeV)} \; = \; 
\begin{array}{lcl}
171 \pm 15 \; \mbox{MeV} \; \; \; (\cite{CAD3})\\[.3cm]
178 \pm 18 \; \mbox{MeV} \; \; \; (\cite{JAMIN1}).
\end{array}
\end{eqnarray}
A recent reanalysis \cite{COL1}, in exactly the same framework, has uncovered
a sizable systematic uncertainty in the hadronic sector. In fact, it is
claimed in \cite{COL1} that after using the Omn\`{e}s representation to
relate the spectral function to the  $K-\pi$ phase shifts, one should
include a background interfering destructively with the resonances. As a
result of this, the area under the hadronic spectral function is much
smaller than that in \cite{CAD3}-\cite{JAMIN1}, leading to a smaller
value of $m_{s}$, viz. \cite{COL1}
\begin{equation}
m_{s} \; \mbox{(1 GeV)} = 140 \pm 20 \mbox{MeV} \; .
\end{equation}
Systematic uncertainties are also present in the theoretical, i.e. the
QCD, sector. For instance, the perturbative QCD expression of the two-point
function involving the vector or axial-vector divergences is generically
of the form
\begin{eqnarray}
\psi \; (Q^2) \; \propto \; m_{s}^{2} \; (Q^2) \; \left( 1 +
a_{1} \; (\frac {\alpha_{s}(Q^2)}{\pi}) \; + \right. \nonumber\\
a_{2} \; (\frac {\alpha_{s}(Q^2)}{\pi})^{2} \; + 
a_{3} \; (\frac{\alpha_{s}(Q^2)}{\pi})^{3} \; + \nonumber\\ 
b_{1} \; m_{s}^{2} \; (Q^2) \; (1 + c_{1} \frac{\alpha_{s}(Q^2)}{\pi} \; + 
\cdots) 
+\nonumber\\
\left. b_{2} \; m_{s}^{4}(Q^2) \; (1 + c_{2} \; \frac{\alpha_{s}(Q^2)}{\pi}
 \; + \cdots) + \cdots \right) \;.
\end{eqnarray}
Knowing both $m_{s}(Q^{2})$ and $\alpha_{s}(Q^{2})$ to a
given order in perturbation theory, the question is: to expand or not in
powers of the inverse logarithms of $Q^{2}$ appearing in Eq.(7)?. It has
been argued in \cite{CHET1} that one should make full use of the
perturbative expansions of the quark mass and coupling (known presently
to 4-loop order) and thus not expand. A strong argument in favour of this
alternative is that, numerically, the non-expanded expression is far more
stable than the truncated one when going from one order in perturbation
theory to the next. In addition, it was shown in \cite{CHET1} that
logarithmic truncation can lead to large overestimates of radiative
corrections. In any case, repeating the analysis of
\cite{CAD3}-\cite{JAMIN1}, but without expanding leads to a higher value
of $m_{s}$, i.e. \cite{CHET1}
\begin{equation}
m_{s} \; \mbox{(1 GeV)} = 203 \pm 20 \mbox{MeV} \; .
\end{equation}
Combining all the above determinations gives the overall result
\begin{equation}
m_{s} (1 GeV) = 170 \pm 50 MeV \; ,
\end{equation}
which provides a realistic estimate of the underlying uncertainties.\\

Turning to the present determination, we  consider the correlator
\begin{eqnarray*}
 \psi_{5} (q^2) = i \; \int \; d^4 \; x \; e^{i q x} 
\end{eqnarray*}
\begin{equation}
  <0|T(\partial^{\mu} \; A_{\mu}(x) \; \partial^{\nu}
   \; A_{\nu}^{\dagger}(0))|0> \; ,
\end{equation}
where $A_{\mu}(x) = :\bar{s}(x)  \gamma_{\mu}  \gamma_{5} u(x):$, and
$\partial^{\mu} \; A_{\mu}(x) = m_{s} \; :\bar{s}(x)  i  \gamma_{5}
\;  u(x):$.
The QCD expression of this two-point function is known 
\cite{CAD3},\cite{JAMIN1}, \cite{CHET1} at the four-loop level in 
perturbative QCD, and up to dimension
six in the non-perturbative sector. Also, the old problem of mass 
singularities has been satisfactorily solved in \cite{CAD3},\cite{JAMIN1}. 
As a result of this, quark mass corrections are also known up to quartic
order. The QCD expression of the Laplace transform of Eq.(10), i.e. 
\begin{eqnarray*}
\psi_{5}^{''}(M^{2})=\hat{L} \left[ \psi_{5}^{''}(Q^{2}) \right] =
\end{eqnarray*}
\begin{equation}
 \int_{0}^{\infty}e^{-s/M^{2}}\frac{1}{\pi}\; \mbox{Im} \; \psi_{5}(s)
\; ds \; ,
\end{equation}
is given by \cite{CAD4}
\begin{eqnarray*}
\psi_{5}^{''}(M^{2})|_{QCD} = \left[ \bar{m}_{s} (M^{2}) \right]^{2} 
M^{4} \biggl[ \psi_{5(0)}^{''} (M^{2}) \; + \biggr. 
\end{eqnarray*}
\begin{equation} 
\biggl. \frac{\psi_{5(2)}^{''}(M^{2})}{M^{2}} 
+ \frac{\psi_{5(4)}^{''}(M^{2})}{M^{4}} 
+ \frac{\psi_{5(6)}^{''}(M^{2})}{M^{6}} + \cdots \; \biggr] \; ,
\end{equation}
\newpage
where
\begin{eqnarray*}
\psi_{5(0)}^{''}(M^{2}) \equiv \hat{L} \; [\psi_{5(0)}^{''}(Q^{2})] =
\frac{3}{8 \pi^{2}}
\Biggl\{ 1 +
\frac{\bar{\alpha}_{s}(M^{2})}{\pi} \Biggr.
\end{eqnarray*}
\begin{eqnarray*}
\left( \frac{11}{3}  
+ 2 \gamma_{E} \right) + \left( \frac{\bar{\alpha}_{s}(M^{2})}{\pi} 
\right)^{2} \left( \frac{5071}{144} - \frac{35}{2} \; \zeta (3) \right.
\end{eqnarray*}
\begin{eqnarray*}
\left. + \; \frac{17}{4} \; \gamma_{E}^{2} 
 + \frac{139}{6} \; \gamma_{E} - \frac{17}{24} \; \pi^{2} \right)
\end{eqnarray*}
\begin{eqnarray*} 
+ \; \left( \frac{\bar{\alpha}_{s}(M^{2})}{\pi} \right)^{3} \left( - 
\frac{4781}{9} + \frac{1}{6} \; a_{1} - \frac{475}{4} \; \zeta (3) \; \gamma_{E}\right.
\end{eqnarray*}
\vspace{.2cm}
\begin{eqnarray*}
+ \; \frac{823}{6} \;
 \zeta (3) + 
\frac{221}{24} \; \gamma_{E}^{3}
+ \frac{695}{8} \; \gamma_{E}^{2}
- \frac{221}{48} \; \gamma_{E} \; \pi^{2}
\end{eqnarray*}
\begin{equation}
\Biggl. \left. + \; \frac{2720}{9} \; \gamma_{E} -
\frac{695}{48} \; \pi^{2} \right) \Biggr\} \; ,
\end{equation}
\begin{eqnarray*}
 \psi_{5(2)}^{''}(M^{2}) \equiv \hat{L} \left[ \psi_{5(2)}^{''} 
 (Q^{2}) \right] = -
\frac{3}{4 \pi^{2}}
\end{eqnarray*}
\begin{equation}
 \; \left[ \bar{m}_{s} (M^{2}) \right]^{2}
 \left[ 1 + 
\frac{\bar{\alpha}_{s}(M^{2})}{\pi} \left( \frac{16}{3} + 4 \gamma_{E} 
\right) \right] \; ,
\end{equation} 
\vspace{.3cm}
\begin{eqnarray*}
\psi_{5(4)}^{''} (M^{2}) \equiv \hat{L} \left[ \psi_{5(4)}^{''} (Q^{2}) \right] = 
\frac{1}{8} \;
< \frac{\alpha_{s}}{\pi} \; G^{2} >
\end{eqnarray*}
\begin{eqnarray*}
 + \; \frac{1}{2} \; < m_{s} \; \bar{s} s > \;
\left[ 1 + \frac{\bar{\alpha}_{s}}{\pi} \left( \frac{11}{3} + 2 \gamma_{E} 
\right) \right]
\end{eqnarray*}
\begin{eqnarray*}
- \; < m_{s} \; \bar{u} u > \; \left[ 1 + \frac{\bar{\alpha}_{s}}{\pi} \left(
 \frac{14}{3} + 2 \gamma_{E} \right) \right]
\end{eqnarray*}
\begin{eqnarray*}
+ \; \frac{3}{28 \pi^{2}} \; m_{s}^{4} \left[ - \frac{233}{36} - \frac{15}{2} \;
\gamma_{E} \right.
\end{eqnarray*}
\begin{equation}
+ \; 2 \frac{\bar{\alpha}_{s}}{\pi} \left( \frac{37}{9} + 
2 \gamma_{E} 
\right) \left. \left( \frac{\pi}{\bar{\alpha}_{s}} - \frac{53}{24} \right) 
\right] \; ,
\end{equation}
and where $\gamma_{E}$ is Euler's constant,  $\zeta(n)$ is Riemann's zeta
function, $a_{1} = 2795.0778$, all numerical coefficients refer to
three flavours and three colours,
and  we have neglected the up-quark mass everywhere. Given the
uncertainties of the method, plus the size of systematic errors, it is
not justified to keep $m_{u}$ different from zero.
In  line with the discussion at the beginning, and following \cite{CHET1}, 
we shall not expand the QCD expressions in inverse powers of logarithms, 
but rather substitute the numerical values of $\alpha_{s}$ 
and $\bar{m}_{s}$ for a given value of $\Lambda_{QCD}$. The 
dimension-six non-perturbative term has been omitted as it is of no 
numerical importance.
The hadronic spectral function associated with the correlator (10)
is very different from that of the vector divergences. There
is, at present, preliminary information from tau-decays \cite{ALEPH} in a 
kinematical range restricted by the tau-mass. We  reconstruct the
spectral function, including in addition to the kaon-pole its radial
excitations K(1460) and K(1830), normalized at threshold according to
conventional chiral symmetry. 
In addition, we  incorporate the resonant sub-channel  $K^{*}(892)-\pi$, 
which is of numerical importance given the narrow width of the $K^{*}(892)$
(the sub-channel $\rho (770)-K$ is numerically negligible). This chiral
symmetry normalization is of the form \cite{CAD4}
\begin{eqnarray*}
\frac{1}{\pi} \; \mbox{Im} \; \psi_{5}(s)|_{K \pi \pi} =
\end{eqnarray*}
\begin{equation}
\frac{M_{K}^{2}}{2f_{\pi}^{2}} \; \frac{3}{2^{8} \pi^{4}} \;
\frac{I(s)}{s(M_{K}^{2} - s)} \; \theta (s - M_{K}^{2}) \;,
\end{equation}
where
\begin{eqnarray*}
I(s) = \int_{M_{K}^{2}}^{s} \; \frac{du}{u} \; (u - M_{K}^{2}) \; (s - u)
\Biggl\{ (M_{K}^{2} - s) \Biggr.
\end{eqnarray*}
\begin{eqnarray*}
\left[ u - \frac{(s+M_{K}^{2})}{2} \right] 
 - \frac{1}{8u} \; (u^{2} - M_{K}^{4}) \; (s - u)
\end{eqnarray*}
\begin{equation}
\Biggl. + \; \frac{3}{4} \;
(u - M_{K}^{2})^{2} |F_{K^{*}} (u)|^{2} \Biggr\} \; ,
\end{equation}
and
\begin{equation}
|F_{K^{*}} (u)|^{2} = \frac{ \left[ M_{K^{*}}^{2} - M_{K}^{2} \right]^{2} +
M_{K^{*}}^{2} \; \Gamma_{K^{*}}^{2}}
{ (M_{K^{*}}^{2} - u)^{2} + M_{K^{*}}^{2} \; \Gamma_{K^{*}}^{2}} \; . 
\end{equation}
The pion mass has been neglected above, in line with the approximation
$m_{u} = 0$ made in the QCD sector, and in our normalization
$f_{\pi} \simeq 93 MeV$. The complete hadronic spectral function is then
\begin{eqnarray*}
\frac{1}{\pi} \; \mbox{Im} \; \psi_{5}(s)|_{HAD} =
2 f_{K}^{2} M_{K}^{4} \;\delta (s - M_{K}^{2})
\end{eqnarray*}
\begin{eqnarray*}
 + \; \; \frac{1}{\pi} \; \mbox{Im} \; \psi_{5}(s)|_{K \pi \pi} 
\frac{[BW_{1}(s) + \lambda BW_{2}(s)]}{(1 + \lambda)}
\end{eqnarray*}
\begin{equation}
+ \; \; \frac{1}{\pi} \; \mbox{Im} \; \psi_{5}(s)|_{QCD} \theta (s - s_{0})\; ,
\end{equation}
where $f_{K} \simeq 1.2 f_{\pi}$, $\mbox{Im} \; \psi_{5}(s)|_{QCD}$ is
the perturbative QCD spectral function modelling the continuum which
starts at some threshold $s_{0}$, $BW_{1,2}(s)$ are Breit-Wigner forms
for the two kaon radial excitations, normalized to unity at threshold,
and $\lambda$ controls the relative importance of the second radial
excitation. The choice $\lambda \simeq 1$ results in a reasonable
(smaller) weight of the K(1830) relative to the K(1460).\\

We have solved the Laplace transform QCD sum rules using the values:
$<\alpha_{s} G^{2}> \simeq 0.024 \mbox{GeV}^{4}$
, $<\bar{s} s> \simeq <\bar{u} u> = -0.01
\mbox{GeV}^{3}$, and allowing $\Lambda_{QCD}$ and $s_{0}$ to vary in the
range: $\Lambda_{QCD} = 280 - 380 \mbox{MeV}$, and $s_{0}= 4 - 8 
\mbox{GeV}^{2}$. 
The results for $m_{s}(1 \mbox{GeV}^{2})$ are very stable against variations
in the Laplace variable $M^{2}$ over the wide range: 
$M = 1 - 4 \mbox{GeV}^{2}$, as well as against variations in the value of 
$s_{0}$ in the above range. The combined result of this determination is
\begin{equation}
{\bar m}_{s}(1 \mbox{GeV}^{2}) = 155 \pm 25 \mbox{MeV} \;.
\end{equation}
The error given above originates exclusively 
from changes in the relevant parameters, and does not reflect possible
systematic uncertainties from the hadronic sector. These could be large,
as discussed in the introduction.
Our result is consistent with the other determinations in the scalar
channel, Eqs.(5),(6),(8), as well as with the result from combining
the determination of $(m_{u}+m_{d})$ with the current algebra ratio of
strange to non-strange quark masses, Eq.(4). It is also in very good
agreement with  recent lattice QCD results reported at this 
conference \cite{LUB} : 
${\bar m}_{s}(1 \mbox{GeV}^{2}) = 155 \pm 15 \mbox{MeV} \;$.

  \end{document}